# Tensile Forces and Shape Entropy Explain Observed Crista Structure in Mitochondria

M. Ghochani[1], J. D. Nulton[2], P. Salamon[2], T. G. Frey[3], A. Rabinovitch[4], A.R.C. Baljon[1]

[1]Department of Physics, San Diego State University, San Diego, California, USA; [2]Department of Mathematical Sciences, San Diego State University; [3]Department of Biology, San Diego State University; [4]Department of Physics, Ben-Gurion University of the Negev, Beer-Sheva, Israel

**Abstract**
A model is presented from which the observed morphology of the inner mitochondrial membrane can be inferred as minimizing the system's free energy. Besides the usual energetic terms for bending, surface area, and pressure difference, our free energy includes terms for tension that we believe to be exerted by proteins and for an entropic contribution due to many dimensions worth of shapes available at a given energy.

In order to test the model, we measured the structural features of mitochondria in HeLa cells and mouse embryonic fibroblasts using 3D electron tomography. Such tomograms reveal that the inner membrane self-assembles into a complex structure that contains both tubular and flat lamellar crista components. This structure, which contains one matrix compartment, is believed to be essential to the proper functioning of mitochondria as the powerhouse of the cell. We find that tensile forces of the order of 10 pN are required to stabilize a stress-induced coexistence of tubular and flat lamellar cristae phases. The model also predicts $\Delta p = -0.036 \pm 0.004$ atm and $\sigma = 0.09 \pm 0.04$ pN/nm.

## INTRODUCTION

Tomograms of mitochondria in a Hela cell and mouse embryonic fibroblast are shown in Fig. 1. The inner mitochondrial membrane, one continuous surface, can be conceptually divided into an inner boundary membrane (IBM) that lies closely apposed to the outer membrane and the crista membranes comprised of multiple cristae that project into the mitochondrial matrix and connect to the IBM via crista junctions. Although for clarity the IBM and each crista are displayed in different colors, the crista membrane and the IBM comprise one complex surface. Each crista membrane contains a lamellar part and several tubular parts. These tubes connect the lamella to the IBM. Beyond these structural principles, the morphology appears highly varied.

In two previous studies (1,2), we investigated how observed morphologies of restricted portions of the inner mitochondrial membranes can be used to infer thermodynamic information regarding typical configurations. While a more first principles approach (3) is of course preferable, it perforce assumes complete knowledge and understanding of the phenomena. Cellular structures represent an intricate interplay between free energy minimizing configurations and a backdrop of molecular agents and constraints. To untangle the complexities of cellular biophysics, our approach can press on where a first-principles approach is not yet accessible.

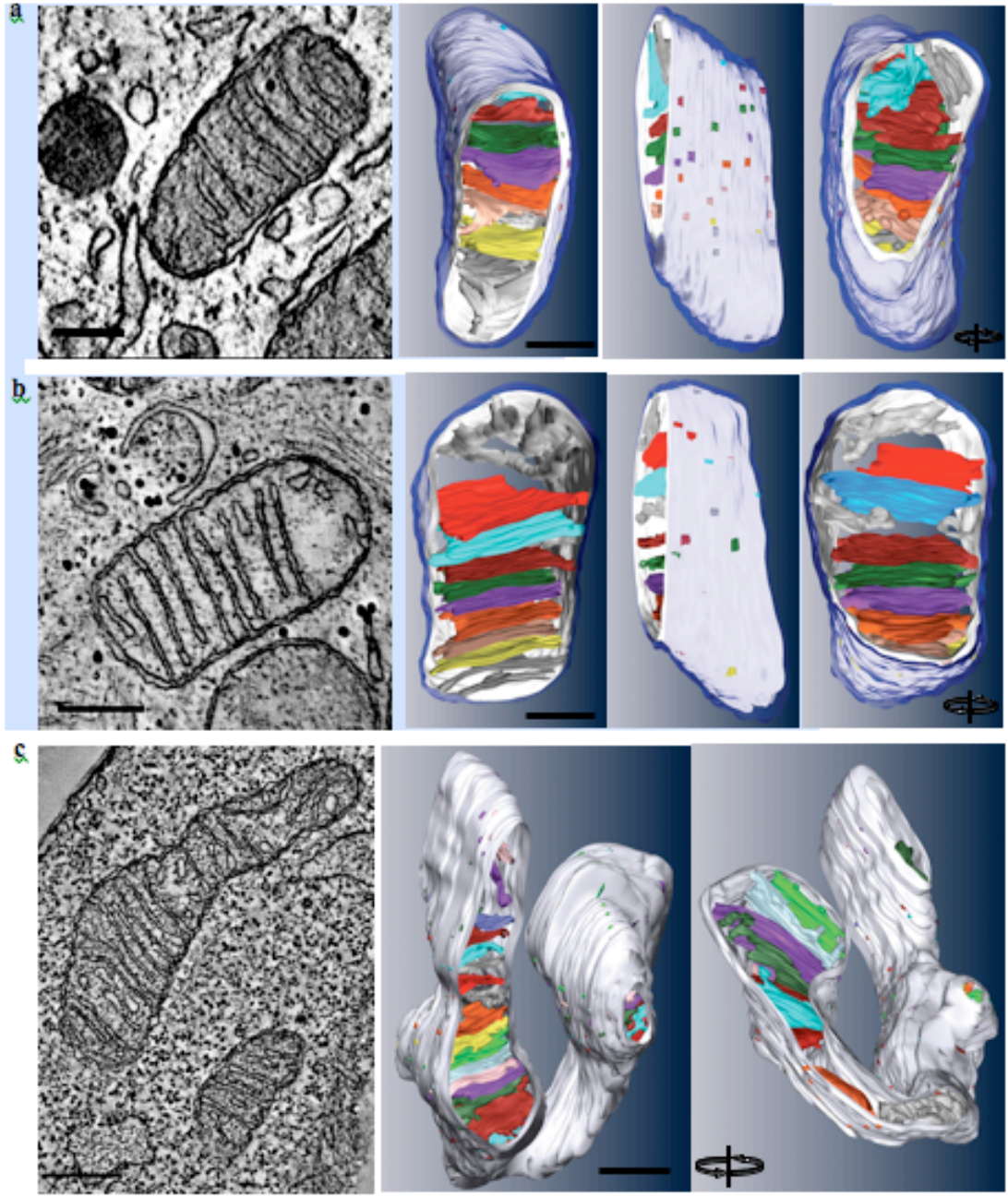

**Figure 1:** Electron tomography of the three mitochondrial volumes. These three volumes were used for measurements that were inserted in the free energy model to obtain thermodynamic parameters in each mitochondrion. On the left, representative sections of constant z from the tomogram volumes are shown and on the right are the 3-D models from volume segmentation and rendering. The successive views rotate about a vertical axis showing different parts of the 3-D models. The outer membrane is shown in translucent blue, the inner membrane in white and cristae in various colors. (a & b) Normal mitochondria from HeLa cells where dual-axis tomography enabled volume reconstruction of the mitochondria. (c) Normal mitochondrion from mouse embryonic fibroblast cells where dual-axis serial tomography on 4 successive sections enabled full volume reconstruction of the mitochondrion. Bar, 250 nm.

Our last study focused on the radii of the tubules and their dependence on the osmotic pressure difference across the inner membrane by assuming that a tubular membrane makes up a thermodynamically stable structure that minimizes free energy. We showed that the observed tubular radii of approximately 10 nm correspond to a pressure that is approximately 0.2 atm lower in the inter membrane space, the compartment between the inner and outer membranes, than in the matrix space, the compartment enclosed by the inner membrane. Our previous model failed however to account for crucial features of the observed morphology. In particular the free energy of the tubular parts was higher than that of the lamellae. Hence the tubes were predicted to be unstable to first order. While some types of mitochondria have only vestigial tubes, many types show the composite structure incorporating both tubes and lamella. In this work we continue the line of reasoning regarding such coexistence by showing that a small tensile force can stabilize the tubular/lamellar structure to first order and calculate the magnitude of this force from observed morphologies. A careful analysis of the stationary states obtained shows that the energy is in fact still not stable to second order. Examining its instability leads to the conclusion that shape entropy plays a significant role in determining the position of the free energy minimum that does not lie very far from the energy minimum.

The paper is organized as follows. In the next section, we set up the simplified model and derive the equations that minimize the energy for such model crista membranes. Then we discuss how we obtain geometrical measurements on the tomograms and use them to deduce values of thermodynamic properties implied by energy minimization: the tensile force required to stabilize the structure and the pressure difference and surface tension of the crista membranes. The discussion that follows introduces the notion that the tubes and lamella can be seen as a two phase equilibrium which leads to many dimensions worth of accessible states at each energy. This admittedly qualitative analysis leads to the conclusion that the observed crista shapes indeed represent a free energy minimum.

**FREE ENERGY MODEL**

The free energy of the inner mitochondrial membrane has a contribution from the inner boundary membrane and each of its crista membranes. In what follows we assume that the free energy is minimized by allowing variations in size of the tubular and lamellar parts of the crista membranes. However we assume that the number of cristae and the number of tubes in each crista remain fixed. Assuming that the shape of the entire inner membrane minimizes free energy, it follows that each of the crista membranes minimizes its free energy. Hence we begin with the free energy for one crista membrane containing one laminar portion and N tubes as shown in Fig 2 and study the conditions under which this free energy is minimized.

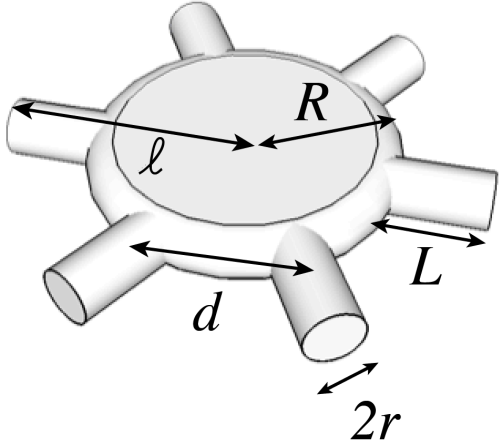

**Figure 2:** The simplified geometry of a crista membrane that is used to set up the equations of the free energy model. The geometry is separate to three different sections: a flat lamellar section at the center, with radius $R$ and thickness $2r$, $N$ tubular membranes, with radius $r$ and length $L$ and a semi cylindrical membrane with radius $r$ that wraps around the edge of the lamellae.

Although a biological membrane is approximately 3 nm thick and contains several different lipids as well as proteins, we treat it as an infinitesimally thin sheet. The energetic part of the free energy can then be described by a Helfrich free energy plus three additional energy terms describing the effect of surface tension, a radial pulling force and a pressure difference across the inner membrane, which separates the inter-membrane region and the matrix space (4-6)

The first term is the Canham-Helfrich curvature energy for an infinitesimally thin fluid sheet, where $\kappa$ is the bending elasticity, $C_1$ and $C_2$ are the principal curvatures and $C_0$ is the spontaneous curvature. The integral is over the surface of the entire crista membrane. The second term is the energetic contribution due to the isotropic surface tension $\sigma$; biomembranes can sustain a strain of a few percent in area before rupturing (7). The expression for the surface energy makes use of the fact that such energy per unit area, i.e., such surface tension, is the intensive parameter conjugate to surface area in the free energy. One may alternatively regard $\sigma$ as the Lagrange multiplier that serves to set the average area of the crista membrane (8). The term $f_{tot}\ell$ represents the effect of the tensile force that pulls on the membrane. Note that the free energy in Eq. 1 only depends on the total volume and area of the membranes in the lamellar and tubular parts. Hence, as we argue below, the configuration shown can be taken as the symmetric representative of a class of energetically equivalent structures. As we shall argue, this accounts for some of the observed variability in the tomograms.

The force on each of the $N$ tubular sections is $f=f_{tot}/N$. $\ell$ is the distance from the center of the structure to the inner boundary membrane.

$$E = \frac{\kappa}{2}\int dA (C_1 + C_2 - C_0)^2 + \sigma \int dA - f_{tot}\ell - \Delta p V \qquad (1)$$

It is equal to the sum of the radius $R$ of flat circular lamellar membranes and the length $L$ of the tubular parts as shown in Fig 2. One may regard $f$ as the Lagrange multiplier that sets the average value of $\ell = R + L$. The last term comes from the difference between the pressure $p_i$ in the inter membrane space (inside the tubular parts) and the pressure $p_m$ in the matrix space $\Delta p = p_i - p_m < 0$.

The free energy of the crista is given by

$$F = E - TS, \quad (2)$$

where T is the ambient temperature and S is the entropy of the membrane. As we will see in the discussion section, this term has important implications for the resulting free energy minimizing shapes. For the present we content ourselves by first examining the internal energy minimizing shapes while neglecting the entropic term. This defines energetic basins in whose vicinity we find the states minimizing free energy.

In order to perform the integrals we separate the crista membrane into three different portions: (*i*) a flat lamellar portion at the center, with radius $R$ and thickness $2r$, (*ii*) $N$ tubular membranes, with radius $r$ and length $L$, and (*iii*) a semi cylindrical membrane with radius $r$ that wraps around the edge of the lamellae. This is an approximation. In particular, we do not model the junctions between the tubes and the lamellae. In our calculations the number of such junctions is always constant. The variations in the size of these junctions are related to the variations in r, which are small. Moreover the energy of the crista junctions is small compared to the total energy. We estimate that these approximations introduce small errors that are on the order of a couple percent.

Next we evaluate the contribution of each of these portions to Eq. 1. Only portions (*ii*) and (*iii*) contribute to the bending energy. For a cylindrical membrane with radius $r$ $C_1=1/r^2$ and $C_2=0$. Since there is no spontaneous curvature $C_0=0$, the bending energy equals (7):
$\frac{\kappa}{2}\int dA(C_1 + C_2 - C_0)^2 = \frac{\kappa}{2}\frac{A_c}{r^2}$, where
$A_c = 2\pi r L N + 2\pi^2 r R$ is the total area of portions (*ii*) and (*iii*).
The second term equals the surface tension times the area of all portions: $\sigma A = \sigma(2\pi R^2 + 2\pi r L N + 2\pi^2 r R)$. The third term in Eq. 1 equals $fN(L+R)$ and the last term:
$\Delta p V = \Delta p(2\pi r R^2 + \pi r^2 L N + \pi^2 r^2 R)$

The energy is minimized by setting its derivatives with respect to the size of the tubules ($r, L$) and the size of the lamellae ($R$), equal to zero. This gives the following set of equations:

$$\frac{1}{r^2}(NL + \pi R) + \left(\frac{\Delta p}{\kappa}\right)(2NLr + 2R^2 + 2\pi r R) - \left(\frac{\sigma}{\kappa}\right)(2NL + 2\pi R) = 0$$

$$\frac{\pi}{r} - \left(\frac{f}{\kappa}\right) - \left(\frac{\Delta p}{\kappa}\right)\pi r^2 + \left(\frac{\sigma}{\kappa}\right)2\pi r = 0 \quad (3)$$

$$\frac{\pi^2}{r} - \left(\frac{f}{\kappa}\right)N - \left(\frac{\Delta p}{\kappa}\right)(4\pi r R + \pi^2 r^2) + \left(\frac{\sigma}{\kappa}\right)(4\pi R + 2\pi^2 r) = 0$$

Eq. 1 has been successfully employed to analyze experiments in which tethers are drawn from large vesicles (3,9,10). When a point force is applied to the vesicle, a thin tether of constant radius forms and grows over time. The stress induces a shape transition and as a result the spherical vesicle coexists with a cylindrical tether. It can be shown that this state is a minimum of the energy in Eq. 1. The pressure difference goes as the inverse radius of the giant vesicle. Since this radius is of the order of a few microns the pressure difference across the membrane can be neglected (3,5). The minimum of the energy in Eq. 1 for tether pulling from large vessicles can be obtained analytically. It follows that $r = \sqrt{\frac{\kappa}{2\sigma}}$ and $f = \frac{2\pi\kappa}{r}$. Both the radius and force are constant during force-induced tether formation. In mitochondria, both the lamellar and tubular parts are at the scale of a few nanometers. Hence the pressure difference is of importance and the full set of Eqs. 3 need to be solved numerically.

**METHODS AND MEASUREMENTS**

In order to solve Eq. 3 for the tensile force $f$, pressure difference $\Delta p$, and surface tension $\sigma$, we need values for the tube radius $r$, the average length of the tubes $L$, the number of tubes $N$, and the size of the lamellar section $R$. These data are obtained from tomograms such as those shown in Fig. 1. In this section we will describe the experimental protocol and the details of the data collection.

**Cell culture and sample preparation for electron microscope tomography**

HeLa cells and mouse embryonic fibroblast cells were cultured and prepared for electron microscopy as described by Sun et al. (11) and Song et al. (12,13), respectively. Up to seven serial semi-thick sections (250-300nm) were cut using a *Diatome* diamond knife on a *Leica EM UC6* ultra-microtome, mounted on 100 mesh copper grids (*Ted Pella 12414-CU*) and poststained with 2% uranyl acetate and Reynolds lead citrate before examination in an FEI Tecnai 12 transmission electron microscope operated at 120 kV. Images were recorded on a 2k x 2k TIETZ 214 CCD camera. Tomograms of single semi-thick sections were determined as described by Sun et al. (11).

**Electron microscope tomography and reconstruction of full mitochondrial tomograms**

Prior to collecting tilt series, optimized positions for aligning the optical and the image axes was found and pre-calibration was done in accordance with Ziese et al. (14,15). Automated tilt series from up to 4 serial sections on a single slot grid were collected using the Xplore 3D tomography suit and corrected for the image shifts and the defocus by use of the pre-calibration and cross-correlation with a comparable image from the previous tilt. Xplore 3D was used to set up band pass filter settings to enhance features of gold markers that do not change much while tilting to get sharp cross-correlation peaks. The grids were then rotated **90°** and a second tilt series from each of the serial sections was recorded. The projection images of each tilt series were processed using the IMOD software suite (16). In the first step the images were aligned to a common axis by choosing a selection of 10 nm fiducial markers from each side of the section and applying a least-squares fit to their positions with TILTALIGN program to find the parameters of alignment equation. The R-weighted backprojection algorithm was used to compute a tomogram for each tilt series. IMOD was used to combine the separate tomograms of each of the two axes using the methods described by Mastronarde (16) and to join tomograms from serial sections (17). Briefly, an approximate transformation was found using the 3-D positions of common fiducial markers in the two alignment procedures. Patches of 64 by 64 by 32 pixels were extracted from each of the volumes, 3-D cross-correlations between them were computed and the results were used in another regression to refine the transformation. After transformation of original tomogram and linear scaling of the densities, the tomograms were combined in Fourier space and the inverse Fourier transform was the final tomogram. Serial combined tomograms were then aligned using linear transformations adjusted manually using visualizations of the alignment between a series of adjacent sections and by a simplex algorithm to find a transformation that minimizes the image difference measures. Tomograms were segmented and rendered as described by Sun et al. (11).

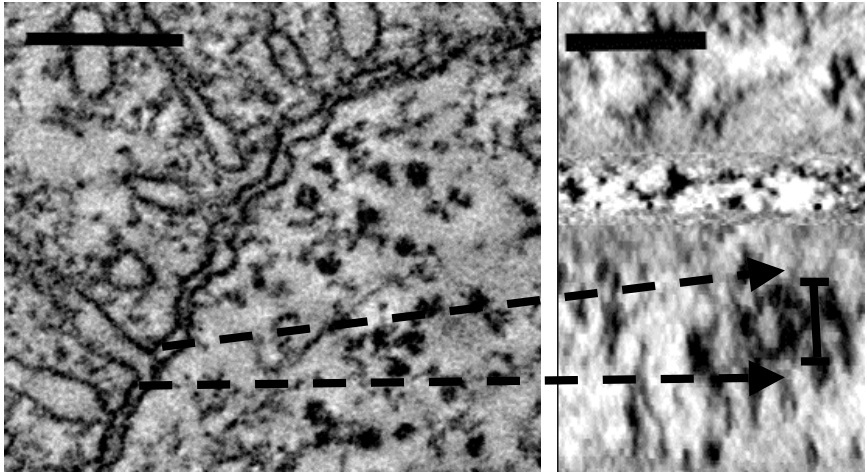

**Figure 3:** (A) The structure of a crista junction through the section of constant z that represents the center of the crista junction from the tomogram volume. (B) The structure of the same crista junction when the stack of the tomogram is reoriented and re-sectioned in the direction normal to the crista junction. The major axis of the crista junction was measured as the vertical dimension through the re-sectioned stack (two-sided line).

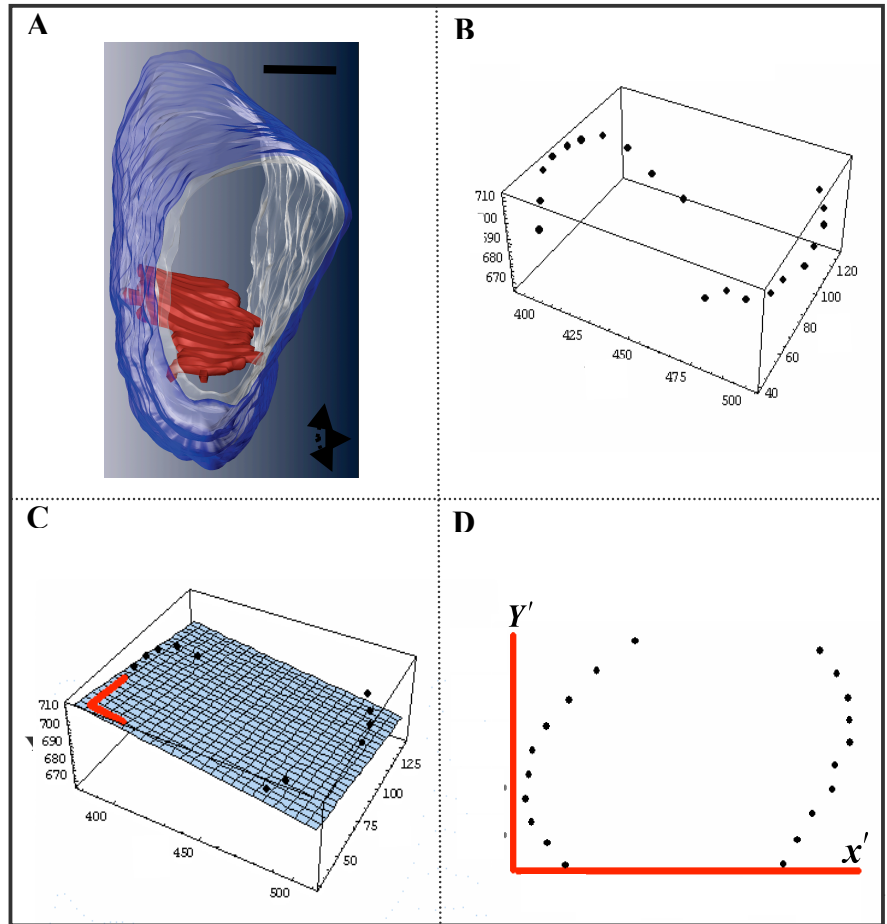

**Figure 4:** (A) A representative crista membrane in the orientation of the tomographic z stack. (B) Boundary points of mitochondrial cross section at the position of the crista membrane in pixel. (C) Best fit regression plane from GRG2 optimization through the boundary points and new set of 2-D basis in the regression plane (red axes). (D) Boundary points in the new 2-D basis from the regression plane in pixel; the best elliptical fit to the 2-D boundary points is obtained by GRG2 optimization method and the size of major and minor axes of the elliptical fit is determined. From these measures the perimeter of the elliptical boundary of mitochondrial cross section at the position of the crista membrane is obtained by solving complete elliptical integral of $1^{st}$ and $2^{nd}$ kinds and by equating this perimeter to that of a circle the distance from the middle of the crista membrane to the IBM ($\ell$) is obtained.

## Measurements

Slices of constant z through dual-axis electron tomograms of two mitochondria from HeLa cells and a full dual-axis electron tomogram of a mitochondrion from mouse embryonic fibroblast cells were analyzed. To this end image slices from tomograms such as those shown in Fig 1 were first smoothed using a Gaussian filter. To improve measurements, contours in each image slice were extracted using Sobel filters which were then combined with the original image slice to get extracted contours in grayscale. Structural features were measured using Image J (18). All measurements were made from the center of the lipid bilayer membranes, which have an average thickness of 7 nm. Crista junction radii were measured by first re-sectioning the tomogram on a plane perpendicular to the junction where the junction joins the IBM as shown in Fig. 3. The direction of re-sectioning was optimized for each junction. Note that the crista junction appears slightly elliptical, however the differences between the major and minor axes were not statistically significant. Thus, we used the average of the major and minor axis of each crista junction as the radii and calculated the average of all crista junctions within a crista to obtain a single radius for each crista.

The length of a tubular part of the crista membrane is the distance between the point where it connects to a lamella and where it connects to the IBM. The point were the tubules connect to lamella can be detected given the sharp change in the direction of the membrane. The value of $L$ was determined by averaging the lengths of all the observed tubes in a single crista.

The distances from the middle of the cristae to the IBM ($\ell$) was determined by fitting an ellipse to a series of 20-30 boundary points on the IBM about each crista within the original tomogram stack (Fig. 4 A and B). The measurement of $\ell$ was used to calculate the radius R of the lamella, which equals $\ell - L$. Fig. 4 C shows how we fit a 2-D plane through the points by minimizing the root mean squares difference between their coordinates and that of the plane. This $X'Y'$ plane is rotated and tilted with respect to the original $XYZ$ axes. In the $X'Y'$ plane the points at which the crista connect to the IBM form an ellipse as can be seen in Fig. 4 D. Complete elliptical integral of $1^{st}$ and $2^{nd}$ kinds were used to calculate the perimeter of the ellipse, and $\ell$ was obtained by equating this perimeter to that of a circle. In order to find the number of tubes ($N$), the center-to-center distances between the tubes were measured to determine the average distance, d, between tubes in a given crista. The best estimate for N is then obtained from N=$2\pi\ell$/d. Note that this value is not an integer. To get another estimate of $N$, the perimeter of the elliptical boundary about each crista was divided by $d$; the value of $N$ that was found by this method was very close to the observed number of tubes in cristae with full volume reconstructions. Using Eqs. 3, the tensile force $f_i$, pressure difference $\Delta p$, and surface tension $\sigma$, have been calculated from the measured values for $\hat{r}_i, \hat{R}_i, \hat{L}_i,$ and $\hat{N}_i$ for each of the 35 cristae studied.

**RESULTS**

These results are shown in Table 1. The bending modulus is taken to be $\kappa = 0.4$ $10^{-19}$ J, a typical value for a biological membrane (7). As can be seen from Eq. 3 the predicted values scale linearly with the bending modulus.
The tensile force is slightly larger than that typical for a molecular motor -10 pN (7). The predicted pressure difference is negative, which is in agreement with the observation that the osmotic pressure in the matrix is larger than in the inter-membrane space (2). The surface tension is within regions of stability for biological membranes, which rupture around 0.2 pN/nm (7). We conclude that the values for tensile force, pressure difference and surface tension predicted by our model are reasonable. For these calculations each crista was treated as a separate thermodynamic system. We expect however that $\Delta p$, representing the pressure difference across the inner membrane would have the same value for all the cristae in equilibrium within the same mitochondrion. Similarly, the surface tension of the membrane σ should have the same value. Our view is that each crista within a given mitochondrion provides us, by way of the procedure described above, with an estimate of the true ambient Δp and σ characterizing the thermodynamic state of the mitochondrion. Our final estimates for Δp and σ (stated below) are computed as mean values accompanied by their standard errors. The actual source of error is manifold and nearly impossible to reliably evaluate. It includes (i) fluctuations within the energetic basin (ii) alterations from the biological state introduced in preparing the cells for electron microscopy, and (iii) errors arising from the manner in which certain of the crista measurements are indirectly inferred from measurements of tomograms. Given these caveats the consistency of our measurements is encouraging.

| $\hat{r}$ (nm) | $\hat{R}$ (nm) | $\hat{L}$ (nm) | $\hat{\ell} = \hat{R} + \hat{L}$ (nm) | $\hat{N}$ | Δp(atm) | σ(pN/nm) | $f$ (pN) |
|---|---|---|---|---|---|---|---|
| 9.9 | 153.6 | 32.2 | 185.8 | 7.6 | -0.053 | -0.023 | 12.9 |
| 10.2 | 174.5 | 36.7 | 211.2 | 15.5 | -0.034 | 0.065 | 17.6 |
| 9.5 | 175.5 | 30.5 | 206.0 | 13.9 | -0.044 | 0.041 | 16.9 |
| 9.5 | 205.0 | 36.4 | 241.4 | 19.7 | -0.030 | 0.10 | 20.2 |
| 10.7 | 225.8 | 40.7 | 266.5 | 20.0 | -0.019 | 0.089 | 18.4 |
| 11.7 | 194.0 | 46.4 | 240.4 | 18.8 | -0.0076 | 0.12 | 19.7 |
| 10.2 | 106.3 | 42.8 | 149.1 | 8.8 | -0.086 | -0.033 | 13.1 |

| $\hat{r}$ (nm) | $\hat{R}$ (nm) | $\hat{L}$ (nm) | $\hat{\ell} = \hat{R} + \hat{L}$ (nm) | $\hat{N}$ | Δp(atm) | σ(pN/nm) | $f$ (pN) |
|---|---|---|---|---|---|---|---|
| 10.5 | 168.1 | 36.7 | 204.8 | 15.1 | -0.031 | 0.064 | 17.3 |
| 9.8 | 174.5 | 31.6 | 206.0 | 14.4 | -0.040 | 0.047 | 17.0 |
| 9.8 | 206.0 | 35.3 | 241.4 | 20.5 | -0.023 | 0.12 | 20.7 |
| 10.7 | 233.2 | 39.6 | 272.8 | 21.8 | -0.014 | 0.11 | 19.6 |
| 11.7 | 195.0 | 45.4 | 240.4 | 18.1 | -0.012 | 0.099 | 18.6 |
| 10.6 | 165.8 | 35.0 | 200.8 | 11.9 | -0.039 | 0.018 | 14.4 |
| 9.9 | 163.6 | 33.6 | 197.2 | 11.7 | -0.047 | 0.015 | 15.1 |
| 9.2 | 147.0 | 27.4 | 174.4 | 9.2 | -0.061 | -0.0081 | 14.8 |

| $\hat{r}$ (nm) | $\hat{R}$ (nm) | $\hat{L}$ (nm) | $\hat{\ell} = \hat{R} + \hat{L}$ (nm) | $\hat{N}$ | Δp(atm) | σ(pN/nm) | $f$ (pN) |
|---|---|---|---|---|---|---|---|
| 9.6 | 199.6 | 45.8 | 245.4 | 16.9 | -0.042 | 0.057 | 17.8 |
| 9.7 | 198.7 | 51.8 | 250.5 | 17.4 | -0.041 | 0.062 | 17.9 |
| 9.1 | 239.5 | 43.4 | 282.9 | 26.7 | -0.017 | 0.17 | 24.2 |
| 9.2 | 252.5 | 55.5 | 308.0 | 30.4 | -0.0011 | 0.23 | 27.1 |
| 10.3 | 256.2 | 48.4 | 304.6 | 25.9 | -0.011 | 0.14 | 21.8 |
| 8.9 | 230.7 | 73.1 | 303.8 | 25.0 | -0.038 | 0.14 | 22.7 |
| 9.1 | 234.9 | 54.5 | 289.4 | 23.5 | -0.034 | 0.12 | 21.3 |
| 8.8 | 166.1 | 49.3 | 215.4 | 15.4 | -0.069 | 0.048 | 18.7 |
| 8.8 | 144.6 | 60.5 | 205.1 | 9.21 | -0.10 | -0.041 | 14.4 |
| 8.9 | 227.7 | 37.9 | 265.6 | 26.6 | -0.015 | 0.19 | 25.3 |
| 9.8 | 227.1 | 48.7 | 275.9 | 28.2 | 0.019 | 0.28 | 29.2 |
| 8.9 | 160.2 | 44.7 | 204.9 | 17.2 | -0.052 | 0.10 | 21.0 |
| 9.3 | 224.1 | 35.8 | 259.9 | 24.3 | -0.017 | 0.16 | 23.2 |
| 10.1 | 201.4 | 45.4 | 246.8 | 22.1 | -0.0096 | 0.16 | 23.1 |
| 9.2 | 152.5 | 42.6 | 195.1 | 18.3 | -0.025 | 0.17 | 23.9 |
| 8.8 | 157.7 | 35.6 | 193.2 | 17.7 | -0.042 | 0.13 | 22.2 |
| 8.9 | 131.2 | 41.4 | 172.7 | 13.9 | -0.072 | 0.062 | 19.4 |
| 8.1 | 133.5 | 37.0 | 170.5 | 18.6 | -0.028 | 0.24 | 28.1 |
| 7.8 | 105.1 | 32.5 | 137.6 | 10.3 | -0.14 | -0.0086 | 18.4 |
| 10.1 | 106.9 | 37.6 | 144.5 | 11.2 | -0.061 | 0.039 | 16.9 |

**Table 1:** Measured values and model predictions for Δp, σ, and $f$ for all cristae. The first table is for HeLa mitochondrion 1, the second for HeLa mitochondrion 2, the third for mouse embryonic fibroblast mitochondria 3.

Our results for the three mitochondria are as follows:

Δp= -0.039 ± 0.010 atm    σ= 0.051 ± 0.023 pN/nm for mitochondrion 1
Δp= -0.033 ± 0.006 atm    σ= 0.058 ± 0.016 pN/nm for mitochondrion 2
Δp= -0.040 ± 0.008 atm    σ= 0.122 ± 0.018 pN/nm for mitochondrion 3

Note that the Δp values are smaller by a factor of 5 than was predicted in the absence of the tensile force (1). To explore the robustness of our parameter values, we also estimated these parameters by finding the minimum distance configuration to our measurements in the space of configurations for the crista network with each crista having the same values for Δp and σ. The values were chosen to minimize the admittedly arbitrary distance

$$D = \sum_{i=1}^{m} \left[ \frac{(\hat{r}_i - r_i)^2}{r_i^2} + \frac{(\hat{R}_i - R_i)^2}{R_i^2} + \frac{(\hat{L}_i - L_i)^2}{L_i^2} + \frac{(\hat{N}_i - N_i)^2}{N_i^2} \right] \quad (4)$$

between the measured values $\hat{r}_i, \hat{R}_i, \hat{L}_i, \hat{N}_i$ and the configuration predicted from a shared value of Δp and σ. The sum is over the m different crista from the same mitochondrion. This method produced the following estimates.

Δp$^*$= -0.030 atm and σ$^*$= 0.068 pN/nm for mitochondrion 1,

Δp$^*$= -0.029 atm and σ$^*$= 0.065 pN/nm for mitochondrion 2,

Δp$^*$= -0.025 atm and σ$^*$= 0.126 pN/nm for mitochondrion 3.

Note that these results are within the standard errors of the first method while exhibiting a systematic overestimate relative to the first method. Nonetheless, the consistency is again encouraging.

Fig. 5 represents another kind of consistency check for our method. It compares the measured values of R, L, and N with some "predicted" values obtained by putting our estimates for the mitochondrial values of Δp and σ back into Eqs. 3. However, these 3 equations contain 7 variables. Therefore, 2

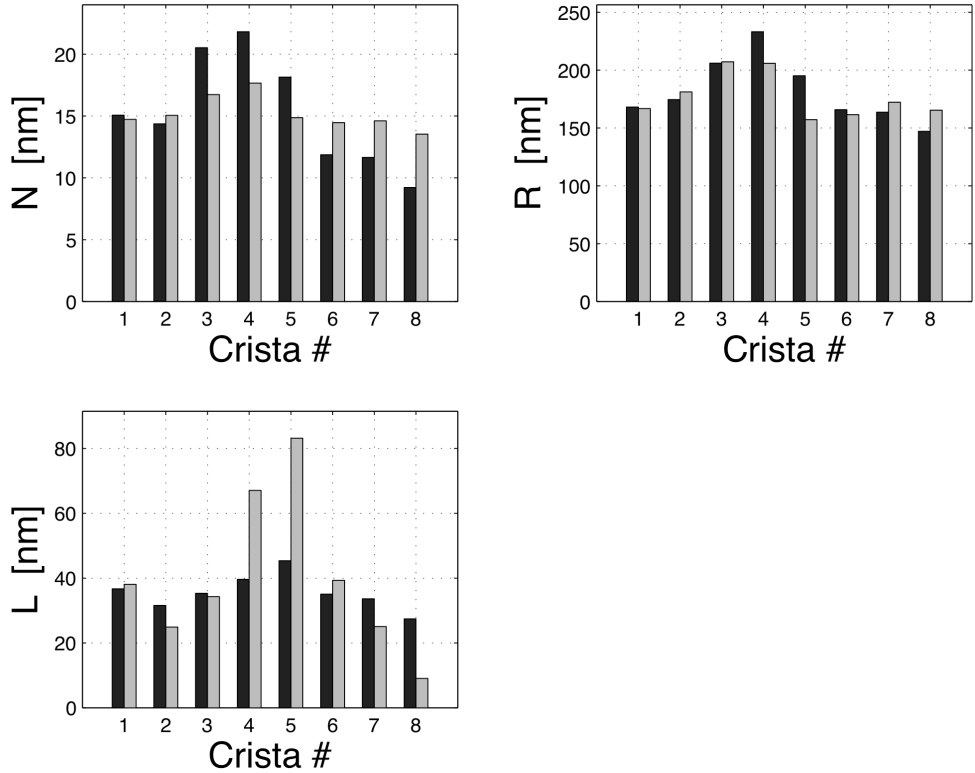

**Figure 5:** Comparison of predicted (gray) and measured (black) values of N, R, and L for the second mitochondria. The predicted values are obtained from the model for $\Delta p$= -0.034 atm $\sigma$= 0.058 pN/nm

additional values must be supplied (beyond the values of $\Delta p$ and $\sigma$) before predictions can be made. We chose to use the measured values for r and $\ell$ for that purpose.

The predicted tensile forces of 15-20 pN are somewhat high compared to the typical force exerted by a molecular motor, which is of the order of 10 pN. One has to keep in mind though that this value depends on our choice of the bending modulus $\kappa$, whose value in the mitochondrial membrane could easily be twice the average value we chose.

We suspect that the tensile force is caused by the dynamin-like GTPase OPA1 that resides in the membrane near the crista junctions. Olichon at al. (21) have shown that the loss of OPA1 perturbs the mitochondrial inner membrane structure. Without this enzyme and the hypothesized tensile force that it exerts, the crista membrane morphology is not stable. Knocking out OPA1 leads to abnormal cristae and the inner membrane fragments into individual vesicular matrix compartments (12). Moreover, knocking out prohibitin results in absence of long isoforms of OPA1 and leads to the vesicular morphology (22). Furthermore, OPA1 is one of the proteins released during programmed cell death (apoptosis) (24), and its loss may be the cause of fragmentation of the inner membrane into vesicular matrix compartments (11).

## DISCUSSION

**The Shape Transition**

The foregoing arguments have shown that adding a tensile force can stabilize the coexistence of tubules and lamellae in model crista. The intuition leading to these arguments came from thinking about the observed coexistence as the likely result of a shape transition under the influence of a tensile force. Such transitions are observed in other materials. For instance, if one pulls on one end of a polymer chain, it separates into a stretched and a coiled part, that coexist at a critical force (19). Likewise, a helical ribbon phase separates under the influence of a force into straight and helical parts (20). It is also similar to what one sees in pulling tethers from vesicles (3). In all these studies the force as a function of extension displays a plateau in the regime where both phases coexist. Extension is made possible by adjusting the fractions of the system in each of the two phases. Along such modes the free energy change is zero and this leads to large fluctuations and critical slowing down.

Similarly, we expect the tubes and lamella of the inner mitochondrial membrane to represent two phases in equilibrium under the applied tensile force. The location of a crista membrane in a mitochondrion dictates its size $\ell$ and hence the relative distribution of the lipids over the tubes and the lamella. Increasing $\ell$ results in lengthening of tubes and shrinking of the lamella. To test this hypothesis, we plot the force $f$ versus $\ell$ in Fig. 6. The curves are obtained from Eq. 3, using the $\Delta p$ and $\sigma$ values of mitochondrion 2. As can be seen, the force depends on both $\ell$ and N at low values of $\ell$, but reaches a plateau value at high extensions. At these extensions the tube radius r is constant.

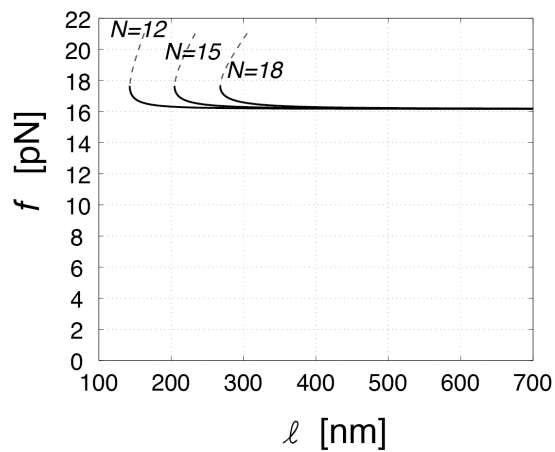

**Figure 6:** Tensile force $f$ as a function of the size of a crista $\ell$. Data are for N=12, 15, and 18. At large $\ell$ the force is independent of both $\ell$ and the number of tubes N.

As can be seen in the second line of Eq. 3, this results in a constant force that is independent of the number of tubes N. While none of the experimental data fall on this plateau, they hover near the transition point for the leveling off – the elbow of the plot.

**Fluctuations**

The results section showed that observed shapes of crista can consistently be associated with real crista shapes and the resulting model can be matched in a consistent way to the
different cristae in the same mitochondrion sharing a pressure difference $\Delta p$ and a surface tension σ. This association succeeds despite the obvious and varied shapes that real cristae assume, which are typically rather unlike our model depicted in Fig. 2. We look for the explanation of such variability by examining the fluctuations.

that we should expect to see in these structures. An examination of the second derivative of the energy of one model crista in the three variables $r_i, R_i, L_i$ reveals an indefinite matrix with two stable directions and one unstable one. Let us examine the cone of instability for one of our examples. It turns out that this cone is very skinny; random perturbations of the stationary state at (r=10.2 nm, R=174.5 nm, L=36.7 nm) resulted in decreased free energy less than 0.5% of the time. As expected, the "null" direction that trades off amount of tube at the cost of lamella is at the boundary of this very skinny cone, i.e. the second derivative along this direction vanishes. This is always the case for a two-phase system along any direction that converts one phase into another (23). The second derivative of the free energy also has a contribution from the TS term and the myriad and varied shapes available to the crista at slightly higher energy E result in an entropic stabilization once one moves some distance exactly along this negative direction.

To see this, we begin by reexamining the energy term in Eq. 1. First we note that for fixed lamellar shape, the energy E only depends on total tube length; moving the tubes around the circumference and lengthening one tube while decreasing another require zero energy. The purely symmetric structure in Fig. 2 is just one representative of a 2n-2 dimensional family of configurations! This degeneracy effectively gives an extra entropic stabilization to the free energy albeit less so for the minimum internal energy configuration than for nearby configurations with some excess tube length and imperfectly circular lamella. With excess tube length and slightly deformed lamella more of the n-1 degrees of freedom that exchange membrane between different tubes can be accessed without deforming the IBM and without changing the total tube length. Increasing the total tube length changes the energy but, for small variations, the entropic contribution more than makes up for it. The amount of phase space volume (and hence entropy) gained by having some play in the total length of the tubes makes a higher dimensional region of phase space accessible. This extra volume of phase space makes a larger contribution to the free energy for states slightly above the energy minimum. Once we are at this higher energy we can also use it to change the lamellar shape slightly away from circular again making accessible a very large additional volume in phase space.

As we argued in the previous subsection, the tubes and lamella represent two phases in equilibrium and thus converting one to the other, i.e., lengthening (say) the tubes and shrinking the lamella must be a variation we can do with $\Delta F = \Delta E - T\Delta S = 0$. In experiments in which tubes are drawn from giant vesicles, the free energy F per unit tube length is zero (3). Moreover, we expect that the shape of the IBM and hence the size of the lamellae (R) and that of the average tube length (L) will fluctuate. As for any equilibrium system immersed in an environment at temperature T, the crista shape will explore variations with free energy changes $\Delta F$ on the order of kT. Given the magnitude of the tensile force f and the surface tension σ, fluctuations with an associated ΔE on the order of kT do not allow for significant changes in the total length of the tubes or in the total surface area of the crista. Equating $f\Delta\ell$ or $\sigma\Delta A$ to kT gives $\Delta\ell$ of about 1 nm and $\Delta A$ about 1 $nm^2$. It is only the entropic contributions that make significant deviations possible.

Such deviations are large as can be seen from the actual shapes observed in tomograms. As argued above, these deviations are at least qualitatively explained by counting the "shape entropy" accessible to the crista. While we believe that it is possible to turn these arguments quantitative by calculating the entropic contributions, such calculations are left for future efforts.

To summarize, our tentative conclusions based on these findings is that the energy minimizing configurations at the given values of Δp, σ, and $f$ adequately represent approximate free energy minima but are in fact only *energetic* stationary points with nearby free energy minima that are stabilized by shape entropy terms and result in the varied structures observed.

**Prolate Spheroid Shapes of Mitochondria**

Healthy mitochondria exhibit elongated shapes. They are attached to molecular motors that pull them along microtubules, which may explain their distortion from the expected spherical shape (24). The tensile forces exerted by the crista membranes on the outer membrane predicted by our model provide an alternative. As can be seen in Table 1 and Fig. 5, these tensile forces are slightly higher in the cristae that reside in the middle of a mitochondrion than in the crista closer to the edges. This is expected since the distortion from a spherical shape is larger in the middle.

The following rough estimate quantitatively support this idea. The deformed mitochondrion has a higher energy than a spherical one. For mechanical equilibrium, the rate of change in energy with $\ell$ should be roughly equal to the sum of all tensile forces. Assuming that the area of the membrane is constant, the increase in the energy is due to an increase in volume and a change in bending energy. The second effect turns out to be negligible. Hence we can write:

$$\Delta p \frac{\partial V}{\partial \ell} = Nf \qquad (5)$$

,where V is the volume between two crista membranes. It is related to the area A of the IBM membrane between these cristae by $V = \frac{A\ell}{2}$. Assuming that the area of the membrane is constant, it follows from Eq. (5) that $A = \frac{2Nf}{\Delta p}$. Let D be the distance between these crista, then

$$D = \frac{A}{2\pi\ell} = \frac{Nf}{\pi\ell\Delta p} \qquad (6)$$

Inserting the data from Table 1 for mitochondria 1, the predicted distance between two cristae equals 96 nm. This compares very well with the average measured value of 82 nm. Moreover, after treatment with etoposide (11) the shape becomes spherical. In this "diseased" state the cristae themselves become vesicular: a topology consistent with the absence of tensile stresses.

**CONCLUSION**

This manuscript takes an important step toward explaining the observed morphology of crista membranes in mitochondria. While such morphology is highly varied, it often consists of coexisting tubes and lamella. Our previous attempt at understanding this coexistence (1) failed; the lamellar portion seems always to have lower free energy. The study described in the present manuscript began with the realization that a tensile force such as the one used in tether pulling experiments, could in fact stabilize the coexistence we were trying to explain. We introduced a simple model incorporating such a tensile force above and found its stationary states. The shapes of real cristae resembled these stationary states sufficiently closely that the parameters of the model could be identified and consistent and reasonable values of the thermodynamic properties ($\Delta p$, $\sigma$, $f$) of the cristae observed could be calculated.

An examination of the second derivative of such internal energy showed there to still exists a narrow cone of instability associated with the degree of freedom that inter-converts lamellar and tubular portions. In this direction, however, we find that energetically degenerate states correspond to a many dimensional space of crista shape parameters and associate this degeneracy with a shape entropy that stabilizes such structures. An admittedly rough count shows the free energy minimum should lie some distance from the energy minimum. Note that the inference of such shape entropy is exactly in line with our overall method – to glean thermodynamic information from observed morphology.

The tubular/lamellar crista structure in normal mitochondria is of functional importance. Healthy mitochondria contain one large matrix compartment in which products of mitochondrial DNA transcription and translation can reach all areas of the inner membrane. Following treatment with etoposide, the tubular and lamellar crista components disappear (11), with the formation of separate vesicular matrix compartments, and impairment of mitochondrial function, which ultimately leads to cell death. During this process, the dynamin-like protein OPA1 is released (25). Hence, we speculate that this motor protein is responsible for the stabilizing tensile forces that are required in our model. We realize that OPA1 could stabilize the observed structures in other ways. One suggestion (26) is that OPA1 is located near the tubular crista and their junctions and controls the structure by wrapping around the tubules. Our model provides an alternative for such theories. Future measurements will hopefully be able to distinguish between these alternative hypotheses.

**Acknowledgment**

We thank Sander Pronk, Karl Heinz Hoffmann and Udo Seifert for useful discussions and suggestions. In addition we thank Dr. Steven Barlow and Sogand Taheri of San Diego State University electron microscopy facility for assistance with preparation of samples for electron microscopy and electron microscope tomography. We acknowledge San Diego State University electron microscopy facility for their support